\pdfoutput=1
\documentclass[12 pt]{article}
\usepackage{graphicx}
\usepackage{amssymb}
\usepackage{amsmath}
\usepackage{amsfonts}
\usepackage{amsthm}
\usepackage[english]{babel}
\usepackage[latin1,ansinew]{inputenc}
\usepackage{a4wide}
\usepackage{color}
\newcommand{\be}{\begin{eqnarray}}
\newcommand{\ee}{\end{eqnarray}}

\begin{document}
\date{}
\title{\textbf{MHD Rankine-Hugoniot jump conditions for shock waves in van der Waals gases}}
\author{\textbf{Raj Kumar Anand} \\Department of Physics, UGC Centre of Advanced Studies
\\University of Allahabad, Prayagraj 211002, India
\\ \textit{email: rkanand@allduniv.ac.in, anand.rajkumar@rediffmail.com}}




\maketitle

%

\begin{abstract}
In this article, we have presented non-relativistic boundary conditions across a magnetohydrodynamic (MHD) shock front propagating in van der Waals gases. The expression for the strength of the non-relativistic MHD shock wave has been obtained, and the Rankine-Hugoniot (R-H) shock jump relations, or boundary conditions, for the pressure, the density, and the particle velocity across an MHD shock front have been derived in terms of a shock compression ratio. The simplified forms of shock jump relations have been written simultaneously for the weak and strong MHD shock waves in terms of the magnetic field strength, the non-idealness parameter, and the ratio of specific heats of the gas. Further, the case of weak shocks has been explored under two distinct conditions, viz., (i) when the applied magnetic field is weak and (ii) when the field is strong, respectively. The case of strong shocks has also been investigated under two distinct ways: (i)  as in the purely non-magnetic case, when the ratio of densities on either side of the shock nearly equals $(\gamma+1)/(\gamma-1)$ or (ii) when the applied magnetic field is large. This is when the ambient magnetic pressure is large as compared with the ambient gas pressure. Finally, the effects on the shock strength and the pressure across the MHD shock front are studied due to the magnetic field strength and the non-idealness parameter of the gases. This study presents an overview of the influence of the magnetic field strength and the non-idealness parameter on the shock strength, the pressure, the density, and the particle velocity across the MHD shock front in van der Waals gases.\\ 


\end{abstract}

\section{Introduction}
\label{intro}
The study of magnetohydrodynamic (MHD) shock waves in non-ideal gases has gained significant attention due to its applications in astrophysics, plasma physics, and high-energy fluid dynamics. The classical Rankine-Hugoniot (R-H) jump conditions, first derived independently by the British engineer William John Macquorn Rankine in 1870 \cite{Rankine1870a, Rankine1870b} and the French ballistician Pierre Henri Hugoniot in 1887 \cite{Hugoniot1887, Hugoniot1889}, describe the conservation laws across shock discontinuities in ideal gases. In shock wave theory, these relations are known as the R-H boundary conditions or R-H relations \cite{Liepmann1957,Whitham1974,Zeldovich2002,Krehl2015}. Shock wave theory has been instrumental in modeling diverse phenomena, from atomic explosions in Earth's atmosphere to supernovae and active galactic nuclei.\\

Magnetohydrodynamic (MHD) phenomena play a crucial role in numerous astrophysical and terrestrial processes, including magnetized stellar winds, planetary nebulae dynamics, synchrotron radiation from supernova remnants, gamma-ray bursts, and dynamo effects in stellar and galactic systems \cite{ Moreau1990, Biskamp2003} The magnetic field strengths encountered in astrophysics range from $10^{-6}$ G in the hot ISM to $10^{12}$ G on the surface of a neutron star. The magnetic fields have a significant effect on the dynamics of astrophysical fluids  \cite{Inoue2009}. The study of magnetogasdynamics has broad applications, spanning plasma physics, nuclear science, and engineering, making it a key area of interdisciplinary research \cite{Freidberg2008}. Applications also extend to industrial processes, including turbulence control in steel casting, drag reduction in duct flows, and fusion reactor design \cite{Sutton2006}.  MHD shock waves have been extensively studied in gas dynamics \cite{Hoffmann1950,Whitham1958,Murray1961}.\\

As shock strength increases, non-ideal gas effects become significant, leading to richer dynamics than those predicted by ideal gas theory. However, under extreme conditions such as high pressures, low temperatures, or strong magnetic fields, the ideal gas assumption breaks down, necessitating more sophisticated equations of state (EoS). The non-ideal gas EoS provides a more accurate description of such non-ideal gases by accounting for finite molecular volume and intermolecular forces. In 1972 Anisimov and Spiner \cite{Anisimov1972} investigated point explosions in low-density non-ideal gases using a simplified equation of state \cite{Landau1958} that accurately describes the medium's behavior. In 1996, Robert and Wu \cite{Roberts1996} also studied intense spherical implosions in both ideal and van der Waals gases, exploring their relevance to inertial confinement fusion and sonoluminescence. In 1873 the van der Waals EoS was derived by Johannes Diderik van der Waals \cite{Waals1873} from kinetic theory considerations, taking into account to a first approximation the size of a molecule and the cohesive forces between molecules. In 1910 Johannes Diderik van der Waals received the Nobel Prize for his work on the EoS for gases and liquids \cite{Waals1967}. Recent advancements in shock wave theory have emphasized the importance of non-ideal gas effects in both weak and strong shock regimes. Wu  and Roberts \cite{Wu1996} studied the structure and stability of a spherical shock wave in a van der Waals gas. Zhao et al. \cite{Zhao2011} showed that shock waves in van der Waals gases exhibit richer thermodynamic and kinematic behaviors compared to ideal gases, particularly in high-energy-density environments. The R-H relations across weak and strong shocks in van der Waals gases was obtained by Anand \cite{Anand2017}.\\

The interaction of MHD shocks with non-ideal gases introduces additional complexity due to the coupling between fluid dynamics and electromagnetic fields. In 1990, Wu \cite{Wu1990} investigated the formation and stability of intermediate shocks in dissipative MHD, highlighting the role of magnetic fields in modifying shock structure. Later, Anand \cite{Anand2013} derived generalized MHD shock jump conditions for non-ideal gases, demonstrating that shock strength and post-shock thermodynamic variables are significantly influenced by both the non-idealness parameter and the applied magnetic field. These studies underscore the need for a refined theoretical framework to describe MHD shocks in van der Waals  gases, particularly in scenarios such as supernova remnants, interstellar medium dynamics, and laboratory plasma experiments.\\ 

Despite these developments, a comprehensive analysis of the MHD R-H jump conditions for van der Waals gases, particularly focusing on the interplay between magnetic fields, non-ideal gas effects, and shock strength, remains an open area of research. \\

This article aims to establish the generalized MHD R-H jump conditions for a van der Waals gas and analyze how the non-idealness parameter and magnetic field strength influence shock dynamics. The findings will contribute to a deeper understanding of MHD shocks in non-ideal plasmas, with implications for astrophysical phenomena, inertial confinement fusion, and high-speed aerodynamics. \\

The structure of this paper is organized as follows: In the next Sect. \ref{conditions} we present the construction of R-H conditions for MHD shock waves. Section \ref{results} contains the analysis with discussion on important components. Concluding remarks are given in Sect. \ref{concl}.

\section{R-H jump conditions across an MHD shock front}
\label{conditions}

The ideal gas EoS was derived based on two key assumptions: (i) the size of gas molecules is negligible relative to the total volume and the spacing between them, and (ii) no intermolecular forces (attractive or repulsive) exist between the molecules. In 1873, van der Waals \cite{Waals1873} developed an empirical EoS for non-ideal gases by addressing these assumptions. To account for the first assumption, he recognized that at high pressures, gas molecules occupy a substantial portion of the total volume. As a correction, he introduced the parameter $b$, representing the excluded volume of the molecules, and subtracted it from the actual molar volume $V$ in the ideal gas EoS: $p=\frac{R T}{V}$, to give $p=\frac{R T}{V-b}$. To remove the second assumption, van der Waals introduced a correction term, represented as $\frac{a}{V^2}$, into the equation to account for intermolecular attractive forces. Mathematically, van der Waals expressed this \cite{Waals1967} as: 
\begin{equation}\label{eq-1}
    p=\frac{R T}{V-b}-\frac{a}{V^2}
\end{equation}
were $R$ is the gas constant, $p$, $V$ and $T$ are the pressure, specific volume, and absolute temperature of the non-ideal gas, respectively, and $a$, $b$, are material-dependent characteristic parameters. The two parameters $a$ and $b$ are constants and, respectively, represent a measure for the attraction between the constituent particles and the effective volume of each particle. For $a\approx b = 0$, the van der Waals EoS reduces to the characteristic equation of an ideal gas $pV = RT$. At high pressures $p$, the term $\frac{a}{V^2}$ is very small relative to $p$ and can be neglected. The term $b$ represents the effect arising from the finite size of the molecules and at low pressures one may omit $b$ from the term containing the specific volume. Conversely, at very high pressures $p\rightarrow \infty$ , the volume $V$ decreases significantly, approaching the value of $b$, which corresponds to the actual volume of the molecules, as mathematically expressed by $\lim_{p \to \infty} V(p) = b$. The fact that only two new constants are involved makes the van der Waals equation relatively easy to use.

The shock wave is a single unsteady wave front with no thickness or a single steady wave of finite thickness. The thickness of shock front is about $10^{-7}$ \textit{m} in air at ambient conditions, and they arise due to the deposition of large amounts of energy in a very small region over short intervals. Since shock waves involve high pressures and temperatures, the van der Waals EoS can be expressed as: $p = \frac{\rho R T}{1 - b \rho}$, where $\rho = 1/V$ is the gas density, and $b$ represents the van der Waals excluded volume, determined by molecular interaction potentials in high-temperature gases. This parameter imposes an upper density limit $\rho_{\text{max}} = 1/b$. Notably, when $b = 0$, the equation reduces to that of an ideal gas (where intermolecular interactions are negligible). This form of the EoS is particularly useful under high-temperature and high-pressure conditions, where the excluded volume dominates over intermolecular forces. The gas constant $R$ and temperature $T$ follow the thermodynamic relation $R = c_p - c_v$, where $c_v = R/(\gamma - 1)$ is the specific heat at constant volume, $c_p = R \gamma / (\gamma - 1)$ is the specific heat at constant pressure, and $\gamma$ is the specific heat ratio ($c_p/c_v$). Following Wu and Roberts \cite{Wu1996}, the internal energy $e$ per unit mass for a non-ideal gas is given by:  $e = c_v T = p \frac{(1 - b \rho)}{\rho (\gamma - 1)}$.

This study incorporates the following assumptions and approximations: (i) The gas follows the van der Waals EoS, (ii) the gas is homogeneous, isotropic, and chemically inert, (iii) viscous and thermal conduction effects are negligible, (iv) the specific heat ratio remains constant and does not vary with temperature, (v) molecular dissociation and ionization are insignificant, (vi) the shock wave moves steadily with an instantaneous step wave profile, (vii) thermal radiation from shock heating is negligible, and (viii) the gas flow is assumed to be one-dimensional. It is important to note that a steady shock with a step wave profile implies an instantaneous transition (zero-rise time) and constant shock velocity, particle velocity (kinematic properties), as well as uniform pressure, density, and internal energy (thermodynamic properties) behind the shock front.

The R-H boundary conditions are a set of equations relating the state variables of the shocked medium to the ones of the undisturbed medium. If the magnetic field is perpendicular to the shock front, then the flow is entirely along the magnetic field lines and remains unaffected by the magnetic field. Therefore, in such a case, the R-H conditions are the same as in the non-magnetic case \cite{Anand2013}. If the magnetic field is parallel to the shock front, then we need to include the magnetic terms in the momentum equation and the energy equation. Here it is assumed that the magnetic field is uniform in the upstream region. 

If $r=R(t)$ be the position of the shock front then the velocity of the shock front is given by $U= \frac{dR}{dt}$. The R-H conditions are obtained from the following principles of conservation of mass, magnetic flux, momentum and energy: 
\begin{eqnarray}\label{eq-2}
\rho(U-u)=\rho_o U, H(U-u)=H_o U,\nonumber \\
p+\frac{\mu H^{2}}{2}+\rho(U-u)^2=p_o+\frac{\mu H_{o}^{2}}{2}+\rho_o U^2,\\
e+\frac{p}{\rho}+\frac{(U-u)^2}{2}+\frac{\mu H^2}{\rho}=e_o+\frac{p_o}{\rho_o}+\frac{U^2}{2}+\frac{\mu H_o^2}{\rho_o}.\nonumber
\end{eqnarray}
where $u$ is the particle velocity, $\rho$ is the density, $p$ is the pressure, $\mu$ is the magnetic permeability, $H$ is the magnetic field, $e$ is the internal energy of the gas per unit mass, The quantities without suffix, i.e., $u, \rho, p$, etc.,  and with suffix, i.e., $u_o, \rho_o, p_o$, etc., denote the values of the quantities behind and in front of the shock front,respectively, in the equilibrium state. The shock jump conditions (\ref{eq-2}) are also valid for curved shock front (e.g., in a spherical medium) because the thickness of the shock front is almost always negligible compared to its radius of curvature.\\
In terms of a shock compression ratio $\rho/\rho_o=\xi$ (say), the  equation (\ref{eq-2}) representing MHD shock conditions may be expressed as: 
\begin{eqnarray}\label{eq-3}
\rho&=&\rho_o \xi, H=H_o \xi, u= U(\xi-1)/\xi, \nonumber\\
U^2&=&\frac{2\xi}{(\gamma+1)-(\gamma-1+2b\rho_o)\xi}\left[(1-b\rho_o)a_o^2 \nonumber \right.\\ && \left.+ \lbrace\gamma+[2-\gamma-b\rho_o(\xi+1)]\xi\rbrace\frac{b_o^2}{2}\right],\\
p&=&p_o+\frac{2\rho_o(\xi-1)}{(\gamma+1)-(\gamma-1+2b\rho_o)\xi}\left[(1-b\rho_o)a_o^2 + (\gamma-1)(\xi-1)^2\frac{b_o^2}{4}\right].\nonumber
\end{eqnarray}
where $a_o=\sqrt{\gamma p_o/\rho_o(1-b\rho_o)}$ is the speed of sound in the unshocked gas, and $b_o=\sqrt{\mu H_o^2/\rho_o}$ is the Alfven speed. The above shock jump conditions (\ref{eq-3}) are in agreement with the well-known MHD shock conditions  for perfect gas \cite{Zeldovich2002,Whitham1958,Murray1961}. The magnetic field strength \cite{Anand2013} is given by the ratio of the Alfven speed to the local speed of sound in the gas, i.e., $b_o^2/a_o^2=\beta^2$ (say). 

\subsection{Weak MHD shock waves}
In the limiting case of weak shocks, $p/p_o $ is very small and the parameter $\xi$ is slightly greater than unity. Therefore, we may write $\xi=1+\varepsilon$, where $\varepsilon(r)$ is another parameter which is negligible in comparison with unity, i.e., $ \varepsilon\ll1$. Under these limitations, equation (\ref{eq-3}) may be written as:
\begin{eqnarray}\label{eq-4}
\rho&=&\rho_o(1+\varepsilon), H=H_o(1+\varepsilon), u= U\varepsilon, \nonumber\\
U^2&=&\frac{2(1-b\rho_o)+(\gamma+1)\varepsilon}{2(1-b\rho_o)}a_o^2 + \frac{2+3\varepsilon}{2}b_o^2,\\
p&=&p_o + \rho_o \varepsilon \left[\frac{2(1-b\rho_o)+(\gamma-1 + 2b\rho_o)\varepsilon}{2(1-b\rho_o)}a_o^2 + \frac{(\gamma-1)\varepsilon^3}{4(1-b\rho_o)}b_o^2\right]\nonumber
\end{eqnarray}

Now we consider the two cases of weak and strong magnetic fields:

\textbf{Case I:} For a weak magnetic field $\beta^2<1$, i.e., when $b_o^2 \ll a_o^2$, or $\mu H_o^2(1-b\rho_o) \ll \gamma p_o$, under this condition the boundary conditions (\ref{eq-4}) for weak shock reduce to 
\begin{eqnarray}\label{eq-5}
\rho&=&\rho_o(1+\varepsilon), H=H_o(1+\varepsilon), u= a_o\varepsilon, \nonumber\\
\frac{U}{a_o}&=& 1+\frac{\lbrace1+\gamma + 3(1-b\rho_o)\beta^2\rbrace\varepsilon}{4(1-b\rho_o)},\quad \\            
\frac{p}{p_o}&=& 1+\frac{\gamma}{1-b\rho_o}\varepsilon.\nonumber
\end{eqnarray}
Equation (\ref{eq-5}) represents the handy forms of R-H jump conditions for weak shocks in the presence of a weak magnetic field.

\textbf{Case II:} For a strong magnetic field $\beta^2>1$, i.e., when $b_o^2 \gg a_o^2$, or $\mu H_o^2(1-b\rho_o) \gg \gamma p_o$, then the boundary conditions (\ref{eq-4}) become 
\begin{eqnarray}\label{eq-6}
\rho&=&\rho_o(1+\varepsilon), H=H_o(1+\varepsilon), u= a_o\varepsilon, \nonumber\\
\frac{U}{a_o}&=& \left[1+\frac{\lbrace1+\gamma+3(1-b\rho_o)\beta^2\rbrace\varepsilon}{4(1-b\rho_o)\beta^2}\right]\beta,\\
\frac{p}{p_o}&=& 1+\frac{\gamma}{1-b\rho_o}\varepsilon.\nonumber
\end{eqnarray}
Equation (\ref{eq-6}) represents the handy forms of R-H conditions for weak shocks in the presence of a strong magnetic field.
 
\subsection{Strong MHD shock waves}
Strong shocks, characterized by high values of $\frac{p}{p_o}$, can occur due to two main conditions: either $\xi$ approaches $\frac{\gamma+1}{\gamma-1}$, or $\frac{b_o^2}{a_o^2}$ is very large. The first scenario is common in traditional gas dynamics, but magnetohydrodynamics introduces an additional possibility that the strong shocks can also form when the magnetic field is extremely intense, regardless of $\xi$ as long as $\xi>1$. Next, we will examine the cases of both weak and strong magnetic fields.

\textbf{Case I:} For a weak magnetic field where $\beta^2 < 1$, which occurs when $b_o^2$ is much smaller than $a_o^2$, or equivalently when $\mu H_o^2(1 - b\rho_o)$ is much less than $\gamma p_o$, the boundary conditions given in (\ref{eq-3}) simplify to 
\begin{eqnarray}\label{eq-7}
\rho&=&\rho_o\xi,  H = H_o\xi,  u = U(\xi-1)/\xi, \nonumber\\
p/p_o&=&1+\left(\chi^{'}a_o^2+A^{'}b_o^2\right)U^2/a_o^4,
\end{eqnarray}
where $\chi^{'}=\frac{\gamma(\xi-1)}{(1-b\rho_o)\xi}$ and $A^{'}=\frac{\gamma(\xi-1)\left[(\gamma-1)(\xi-1)^2-2\lbrace[2-b\rho_o(\xi+1)-\gamma]\xi+\gamma\rbrace\right]}{4(1-b\rho_o)^2\xi}$.

Equation (\ref{eq-7}) provides a convenient formulation of the R-H jump conditions for strong shock waves under the influence of a weak magnetic field.

\textbf{Case II:} In the case of a strong magnetic field where $\beta^2 > 1$, meaning $b_o^2$ is much greater than  $a_o^2$, or equivalently,  $\mu H_o^2(1 - b\rho_o)$ dominates $\gamma p_o$ then the boundary conditions (\ref{eq-3}) simplify to:
\begin{eqnarray}\label{eq-8}
\rho&=&\rho_o\xi, H = H_o\xi, u = U(\xi-1)/\xi, \nonumber\\
p/p_o&=&1+\chi\left(b_o^2+A a_o^2\right)U^2/a_o^2 b_o^2,
\end{eqnarray}
where $\chi=\frac{\gamma(\xi-1)^3(\gamma-1)}{2(1-b\rho_o)\xi\lbrace[2-b\rho_o(\xi+1)-\gamma]\xi+\gamma\rbrace}$ and $A=\frac{4(1-b\rho_o)}{(\gamma-1)(\xi-1)^2}-\frac{2(1-b\rho_o)}{[2-b\rho_o(\xi+1)-\gamma]\xi+\gamma}$.

Equation (\ref{eq-8}) offers a compact representation of the Rankine-Hugoniot conditions applicable to strong shock waves under the influence of a strong magnetic field.

\section{Results and Discussion}
\label{results}

This section presents an analysis of MHD R-H jump conditions derived for one-dimensional MHD shock waves propagating in van der Waals gases. It is worth mentioning that these R-H shock jump conditions are valid for the exploding and imploding MHD shock waves and reduce to the well-known classical MHD R-H conditions \cite{Whitham1958,Murray1961} for the MHD shock waves propagating in an ideal gas when $b\rho_o$, the non-idealness parameter, becomes zero. The magnetic field strength is measured by the ratio of Alfven speed to sound speed in the medium ahead of the shock front, i.e., $b_o^2/a_o^2=\beta^2$ (say). The typical values of parameters are taken as $1<\xi<2.65$, $0<b\rho_o<0.2500$, $0<\beta^2<0.1<20$, and $\gamma =1.4$ for numerical computation of shock strength and pressure using Mathematica-8. Obviously, the value $\beta^2=0$ corresponds to a non-magnetic case \cite{Anand2013}; however, $\beta^2>0$ corresponds to a magnetic case. The influence of weak and strong magnetic fields on the shock strength and the pressure across the shock front has been investigated, respectively, for the weak and the strong MHD shock waves in van der Waals gases.

\subsection{Weak MHD shock waves}
Now we explore the influence of the non-idealness parameter and the magnetic field strength on the weak shock waves propagating in van der Waals gases under two conditions viz., (i) when the magnetic field is weak and (ii) when the field is strong, respectively.

\begin{figure}   
   \begin{minipage}{0.4\textwidth}
     \centering
     \includegraphics[width=1.2\linewidth, trim=0 .1cm .1cm 0, clip]{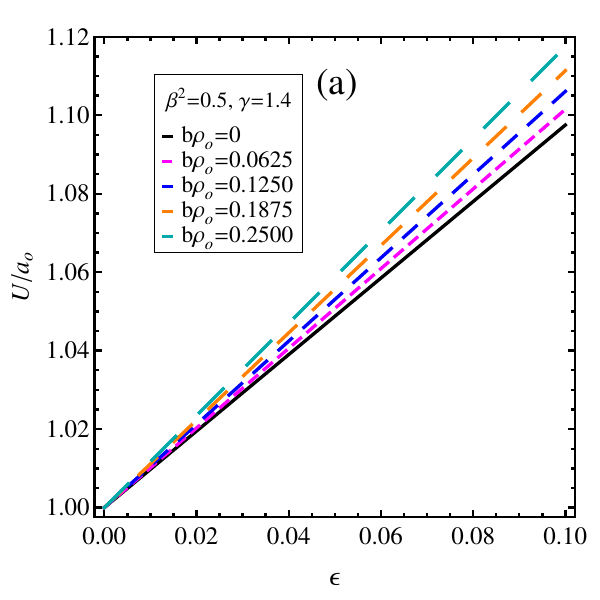}
       \end{minipage}\hfill
   \begin{minipage}{0.4\textwidth}
     \centering
     \includegraphics[width=1.2\linewidth, trim=0 .1cm .1cm 0, clip]{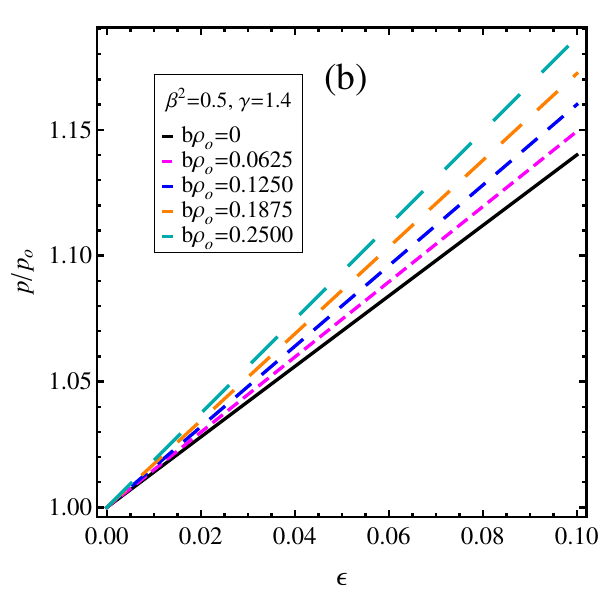}
     \end{minipage}
     \begin{minipage}{0.4\textwidth}
     \centering
     \includegraphics[width=1.2\linewidth, trim=0 .1cm .1cm 0, clip]{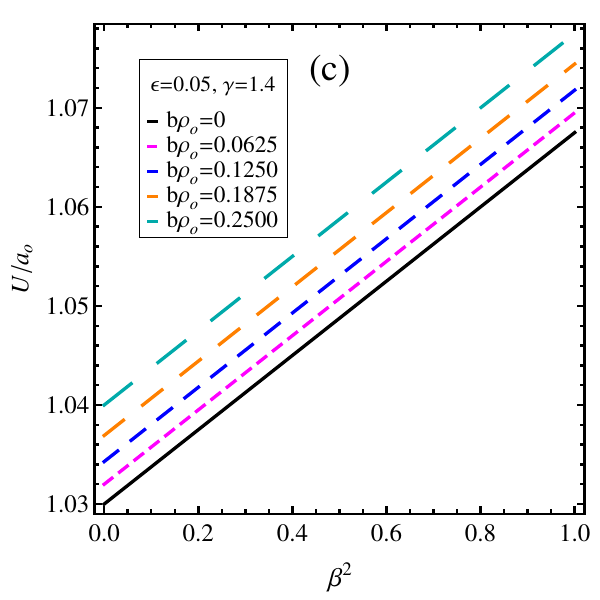}
       \end{minipage}\hfill
   \begin{minipage}{0.4\textwidth}
     \centering
     \includegraphics[width=1.2\linewidth, trim=0 .1cm .1cm 0, clip]{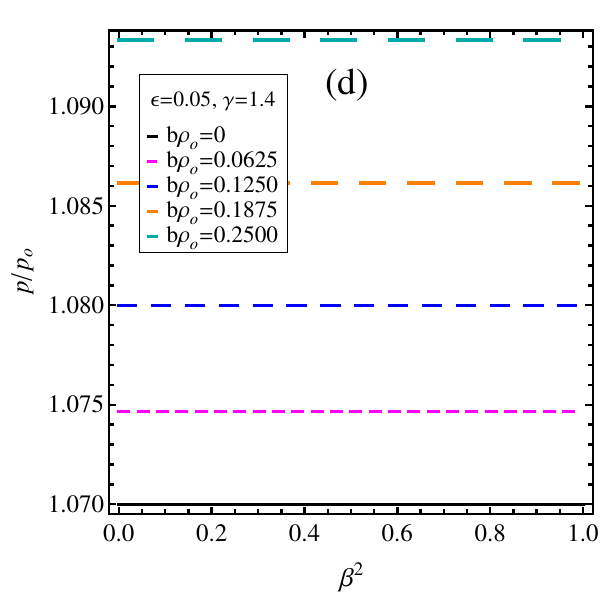}
     \end{minipage}
     \caption{The profiles of (a) $U/a_{o}$ vs. $\varepsilon$, (b) $p/p_{o}$ vs. $\varepsilon $, (c) $U/a_{o}$ vs. $\beta^2$,  (d) $p/p_{o}$ vs. $\beta^2$ in case of weak MHD shock waves in the presence of a weak magnetic field.}\label{figure1a_1d}
\end{figure}
\begin{figure}   
   \begin{minipage}{0.4\textwidth}
     \centering
     \includegraphics[width=1.2\linewidth, trim=0 .1cm .1cm 0, clip]{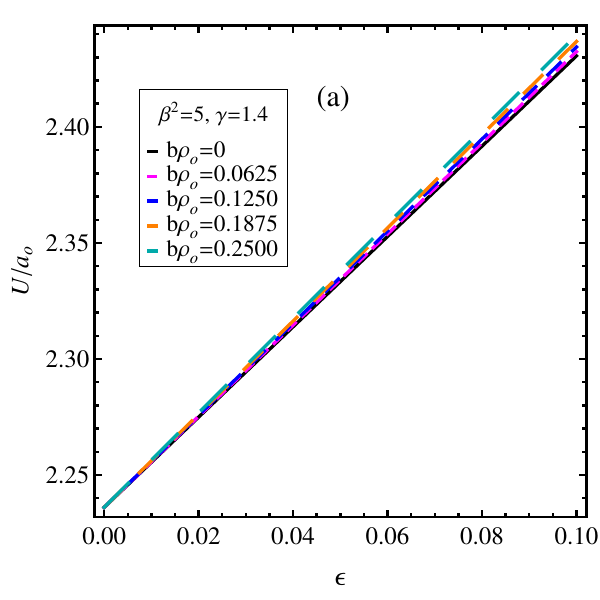}
       \end{minipage}\hfill
   \begin{minipage}{0.4\textwidth}
     \centering
     \includegraphics[width=1.2\linewidth, trim=0 .1cm .1cm 0, clip]{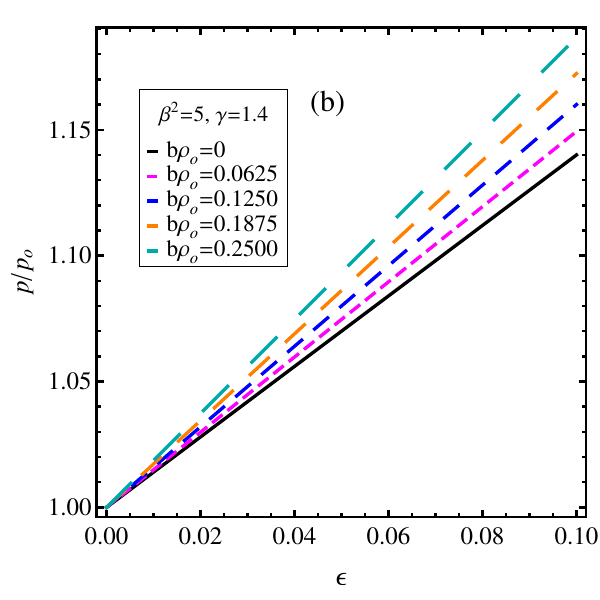}
     \end{minipage}
     \begin{minipage}{0.4\textwidth}
     \centering
     \includegraphics[width=1.2\linewidth, trim=0 .1cm .1cm 0, clip]{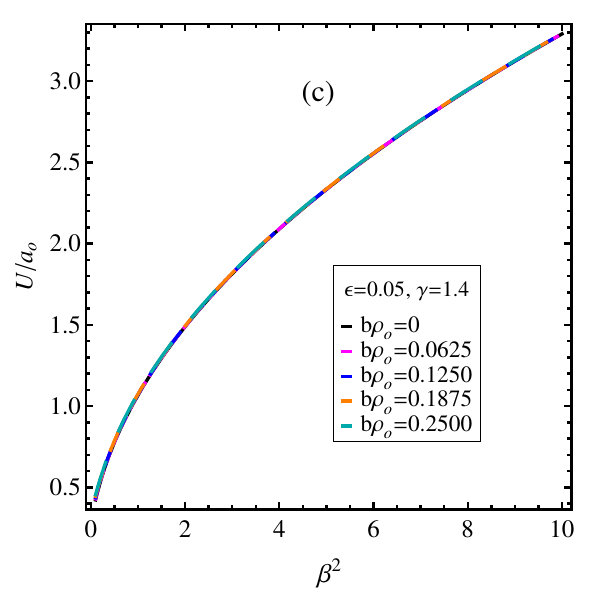}
       \end{minipage}\hfill
   \begin{minipage}{0.4\textwidth}
     \centering
     \includegraphics[width=1.2\linewidth, trim=0 .1cm .1cm 0, clip]{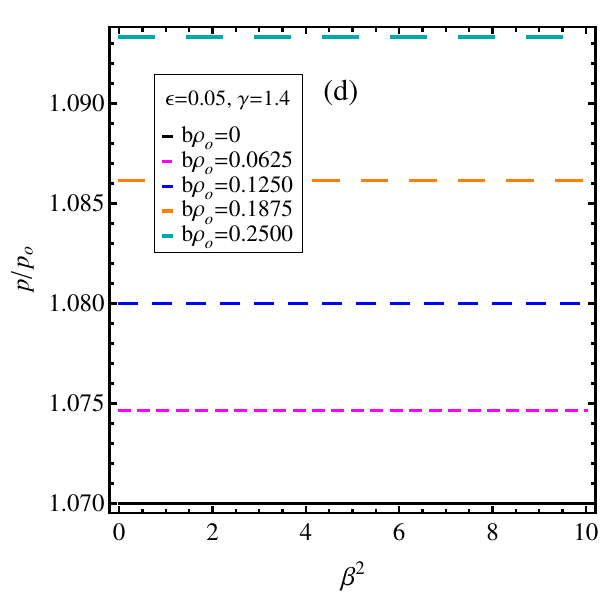}
     \end{minipage}
     \caption{The profiles of (a) $U/a_{o}$ vs. $\varepsilon$, (b) $p/p_{o}$ vs. $\varepsilon$, (c) $U/a_{o}$ vs. $\beta^2$, (d) $p/p_{o}$ vs. $\beta^2$ in case of weak MHD shock waves in the presence of a strong magnetic field.}\label{figure2a_2d}
\end{figure}
 
\textbf{Case I:} For weak magnetic fields, the handy forms of MHD R-H jump conditions for weak shock waves are given by equation (\ref{eq-5}). The shock strength and the state variables, i.e., the pressure, density, and particle velocity, are dependent on a parameter $\varepsilon$, which is negligible in comparison with unity, the magnetic field strength $\beta^2$, and the adiabatic index of gas $\gamma$. The numerical computations of the shock strength and the pressure have been carried out taking the parameters as $0<\varepsilon<0.1$, $0<\beta^2<1$, and $\gamma =1.4$. The variations in the shock strength and the pressure, respectively, with parameter $\epsilon$ and magnetic field strength $\beta^2$ for $\varepsilon=0.05$, $\beta^2=0.5$, $\gamma = 1.4$, and various values of $b\rho_o$ are shown in figure \ref{figure1a_1d}. It is important to note from figure \ref{figure1a_1d} (a)-(b) that the shock strength and the pressure increase with the parameter $\varepsilon$. The shock strength also increases with the non-idealness parameter $b\rho_o$. It is also obvious from figure \ref{figure1a_1d} (c)-(d) that an increase in the magnetic field strength $\beta^2$ leads to an increase in the shock strength. However, the pressure remains almost unchanged with the magnetic field strength.

\textbf{Case II:} For a strong magnetic field, the handy forms of MHD R-H jump relations for weak magnetohydrodynamic shock are given by equation (\ref{eq-6}). The shock strength and the state variables, i.e., the pressure, density, and particle velocity, are dependent on a parameter $\varepsilon$, which is negligible in comparison with unity, the magnetic field strength $\beta^2$, and the adiabatic index of gas $\gamma$. The numerical computations of the shock strength and the pressure have been carried out taking the parameters as $0<\varepsilon<0.1$, $0<\beta^2<10$, and $\gamma =1.4$. The variations in the shock strength and the pressure, respectively, with parameter $\varepsilon$ and magnetic field strength $\beta^2$ for $\varepsilon=0.05$, $\beta^2=5$, $\gamma = 1.4$, and various values of non-idealness parameter $b\rho_o$ are shown in figure \ref{figure2a_2d}. It is obvious from figure \ref{figure2a_2d}(a)-(b) that the shock strength and the pressure increase with the parameter $\varepsilon$. The shock strength and pressure also increase with the non-idealness parameter $b\rho_o$. Figure \ref{figure2a_2d}(c)-(d) shows that an increase in the magnetic field strength $\beta^2$ leads to an increase in the shock strength. However, the pressure remains independent of the magnetic field strength.

\subsection{Strong MHD shock waves} 
Now we investigate the influence of the non-idealness parameter and the magnetic field on the shock strength and the pressure across the strong MHD shock front in van der Waal gases under two conditions, viz., (i) when the magnetic field is weak and (ii) when the field is strong, respectively.

\begin{figure}   
   \begin{minipage}{0.4\textwidth}
     \centering
     \includegraphics[width=1.2\linewidth, trim=0 .1cm .1cm 0, clip]{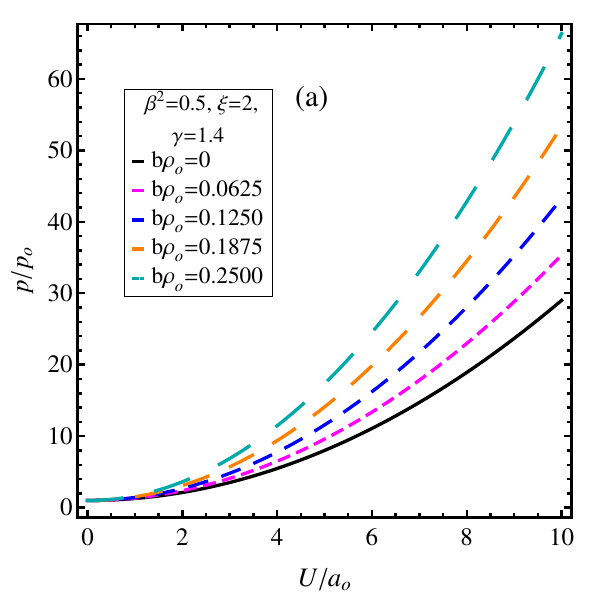}
       \end{minipage}\hfill
   \begin{minipage}{0.4\textwidth}
     \centering
     \includegraphics[width=1.2\linewidth, trim=0 .1cm .1cm 0, clip]{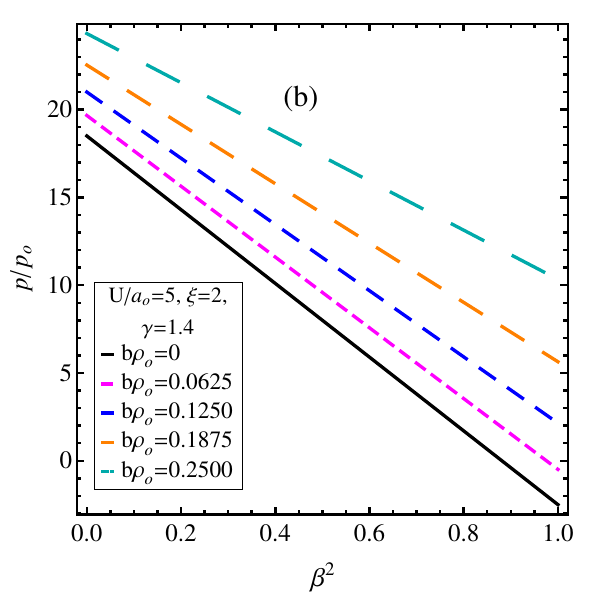}
     \end{minipage}
     \begin{minipage}{0.4\textwidth}
     \centering
     \includegraphics[width=1.2\linewidth, trim=0 .1cm .1cm 0, clip]{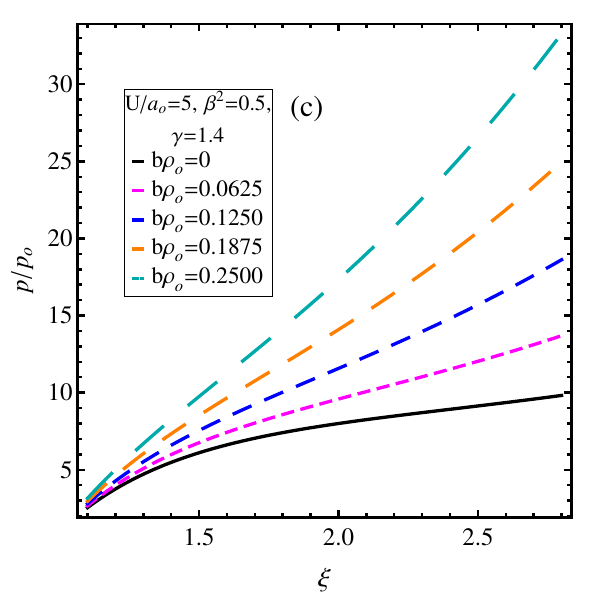}
       \end{minipage}\hfill
       \caption{The profiles of (a) $p/p_{o}$ vs. $U/a_o$, (b) $p/p_{o}$ vs. $\beta^2$, (c) $p/p_{o}$ vs. $\xi$ in case of strong MHD shock waves in the presence of a weak magnetic field.}\label{figure3a_3c}
\end{figure}
\begin{figure}   
   \begin{minipage}{0.4\textwidth}
     \centering
     \includegraphics[width=1.2\linewidth, trim=0 .1cm .1cm 0, clip]{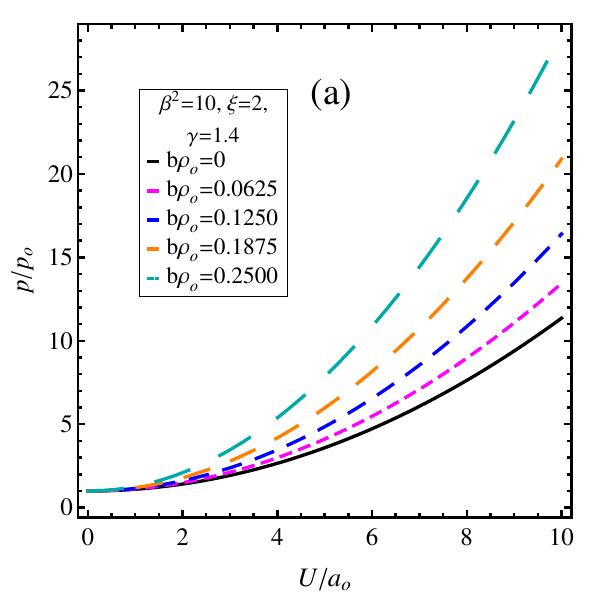}
       \end{minipage}\hfill
   \begin{minipage}{0.4\textwidth}
     \centering
     \includegraphics[width=1.2\linewidth, trim=0 .1cm .1cm 0, clip]{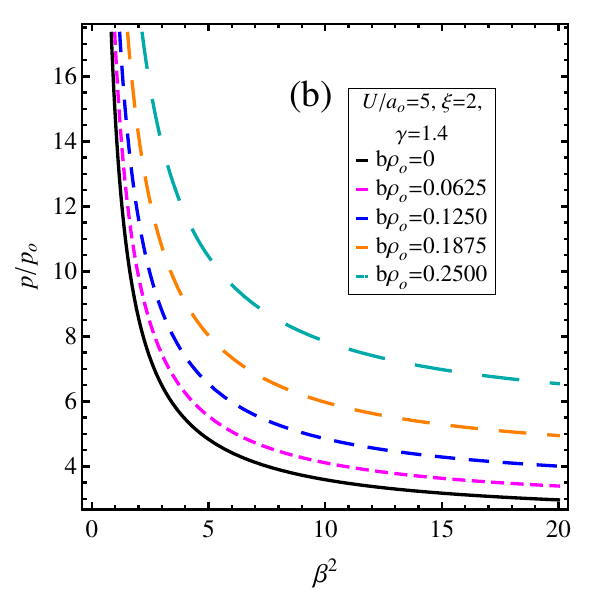}
     \end{minipage}
     \begin{minipage}{0.4\textwidth}
     \centering
     \includegraphics[width=1.2\linewidth, trim=0 .1cm .1cm 0, clip]{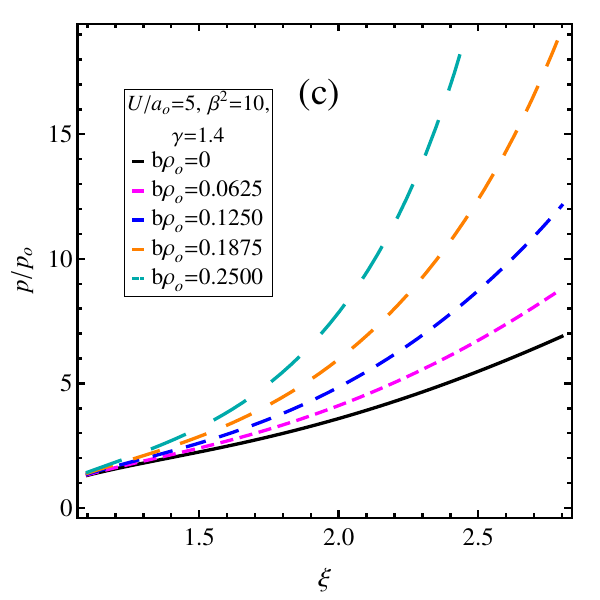}
       \end{minipage}\hfill
       \caption{The profiles of (a) $p/p_{o}$ vs. $U/a_o$, (b) $p/p_{o}$ vs. $\beta^2$, (c) $p/p_{o}$ vs. $\xi$ in case of strong MHD shock waves in the presence of a strong magnetic field.}\label{figure4a_4c}
\end{figure}

\textbf{Case I:} For weak magnetic fields, the handy forms of MHD R-H boundary conditions for strong MHD shock waves are given by equation (\ref{eq-7}). The pressure $p/p_o$ across the shock front is dependent on the shock strength $U/a_o$, the magnetic field strength $\beta^2$, the adiabatic index of gas $\gamma$, and a parameter $\xi$, which is known as the shock compression ratio. The numerical computation of the pressure has been carried out taking the parameters as $0<U/a_o<5$, $0<\beta^2<1$, $1<\xi<2.65$, and $\gamma =1.4$. The variations in the pressure, respectively, with the shock velocity $U/a_o$, the magnetic field strength $\beta^2$, and the shock compression ratio $\xi$ for $\xi=2$, $\beta^2=0.5$, $\gamma = 1.4$, and various values of $b\rho_o$ are shown in figure \ref{figure3a_3c}. The pressure increases with the shock strength and the shock compression ratio; however, it decreases with the magnetic field strength. The pressure also increases with the non-idealness parameter $b\rho_o$.\\
It is notable that in the presence of weak magnetic fields, the trends of variations of the pressure across the strong MHD shock front in van der Waals gases are similar to that of across the strong MHD shock front in an ideal gas.

\textbf{Case II:} For strong magnetic fields, the handy form of R-H conditions for strong MHD shock waves in the presence of a strong magnetic field is given by equation (\ref{eq-8}). The pressure $p/p_o$ across the shock front is dependent on the shock strength $U/a_o$, the magnetic field strength $\beta^2$, the adiabatic index of gas $\gamma$, and a parameter $\xi$, which is known as the shock compression ratio. The numerical computation of the pressure has been carried out taking the parameters as $0<U/a_o<5$, $0<\beta^2<20$, $1<\xi<2.65$, and $\gamma =1.4$. The variations in the pressure, respectively, with the shock strength $U/a_o$, the magnetic field strength $\beta^2$, and the shock compression ratio $\xi$ for $\xi=2$, $\beta^2=10$, $U/a_o=5$, $\gamma = 1.4$, and various values of $b\rho_o$ are shown in figure \ref{figure4a_4c}. The pressure increases with the shock strength $U/a_o$ and the shock compression ratio $\xi$. It is obvious from figure \ref{figure4a_4c}(b) that the pressure first decreases rapidly and then becomes almost constant with the magnetic field strength $\beta^2$. The pressure with the non-idealness parameter $b\rho_o$ also increases.\\
It is notable that in the presence of strong magnetic fields, the trends of variations of the pressure across the strong MHD shock front in van der Waals gases are similar to that of across the strong MHD shock front in an ideal gas.

\section{Conclusions}
\label{concl}
The present study has shown that the shock strength and the pressure across the MHD shock front in van der Waals gases are mainly affected by the non-idealness parameters and the strength of the magnetic field. The strength of weak shock waves increases with the magnetic field strength. However, the pressure across the weak MHD shock front remains almost unchanged with the magnetic field strength. The pressure across a strong MHD shock front increases with the shock strength and the non-idealness parameter of the gas. However, it decreases with the magnetic field strength. The trends of variations of the shock strength and the pressure across the MHD shock front in van der Waals gases are similar to those across the MHD shock front in ideal gas, in general.

\begin{center}
\textbf{APPENDIX}
\end{center}
In the limiting case of a strong shock, $p/p_o$ is large. In the magnetic case this is achieved in two ways:
 
Case I, the purely non-magnetic way when $\xi-(\gamma+1)/(\gamma-1)$ is small.

Case II, when $b_o^2 \gg a_o^2$, i.e., $\mu H_o^2(1-b\rho_o)/\gamma p_o\gg 1$, i.e., the ambient magnetic pressure is very much greater than the gas pressure in equilibrium state. In terms of $\xi$, the shock may be strong (i) for $\xi$ values just greater than one and $\beta^2$ should be large and (ii) for $\xi$ values close to $(\gamma+1)/(\gamma-1)$, corresponds to the case of shock of infinite shock strength. In terms of $\xi$, the boundary conditions (\ref{eq-3}) reduce to
\[ \rho=\rho_o\xi, H=H_o\xi, u= U(\xi-1)/\xi, \]
\[ U^2=\frac{2(1-b\rho_o)\xi a_o^2}{(\gamma+1)-(\gamma-1+2b\rho_o)\xi}\left[1 + \frac{[\gamma+\lbrace(2-\gamma)-b\rho_o(\xi+1)\rbrace\xi]\mu H_o^2}{2\gamma p_o}\right], \]
\[ p/p_o = L + \chi U^2/a_o^2,\]

where $L=\frac{(\gamma+1-2b\rho_o)\xi-(\gamma-1)}{(\gamma+1)-(\gamma-1+2b\rho_o)\xi}$ and $\chi=\frac{\gamma(\gamma-1)(\xi-1)^3}{2\xi\lbrace\gamma+[2-\gamma-b\rho_o(\xi+1)]\xi\rbrace}$.

The above relations are the MHD R-H boundary conditions for strong MHD shock waves. \textbf{Table \ref{tab:1}} shows computed values of $U/a_o$ taking fixed values of $\xi=2$, $\gamma=1.4$, and $b\rho_o = $ 0, 0.03125, 0.06250, 0.09375, 0.12500, 0.15625, 0.18750, 0.21875, 0.25000 for various values of $\beta^2$ from 0 to 20. 

\begin{table}
\caption{MHD strong shock waves: Computed values of $U/a_o$ taking $\xi=2$ and $\gamma=1.4$.}
\label{tab:1}
\begin{tabular}
{p{0.6 cm} p{1.5 cm} p{1.5 cm} p{1.5 cm} p{1.5 cm} p{1.5 cm} p{1.5 cm} p{1.5 cm} p{1.5 cm} p{1.5 cm}}
\hline\noalign{\smallskip}
$U/a_o$&$\leftarrow--$&$----$&$----$&$----$&$-b\rho_o-$&$----$&$----$&$----$&$--\rightarrow$\\
            
$\beta^{2}$& 0 & 0.03125 & 0.06250  &  0.09375 & 0.12500 & 0.15625  & 0.18750 & 0.21875 & 0.25000\\
 \noalign{\smallskip}\hline\noalign{\smallskip}
  0 & 1.58114 & 1.62084 & 1.66667 & 1.72023 & 1.78377 & 1.86052 & 1.95538 & 2.07614 & 2.23607\\
 .1 & 1.68077 & 1.71879 & 1.76278 & 1.81434 & 1.87568 & 1.95002 & 2.04220 & 2.15998 & 2.31661\\
 .2 & 1.77482 & 1.81145 & 1.85392 & 1.90381 & 1.96330 & 2.03558 & 2.12548 & 2.24069 & 2.39444\\
 .3 & 1.86414 & 1.89960 & 1.94079 & 1.98926 & 2.04717 & 2.11769 & 2.20561 & 2.31859 & 2.46982\\
 .4 & 1.94936 & 1.98383 & 2.02393 & 2.07118 & 2.12774 & 2.19673 & 2.28293 & 2.39396 & 2.54296\\
 .5 & 2.03101 & 2.06463 & 2.10379 & 2.14999 & 2.20537 & 2.27303 & 2.35772 & 2.46702 & 2.61406\\
 .6 & 2.10950 & 2.14239 & 2.18072 & 2.22601 & 2.28035 & 2.34685 & 2.43020 & 2.53799 & 2.68328\\
 .7 & 2.18518 & 2.21742 & 2.25504 & 2.29951 & 2.35295 & 2.41841 & 2.50059 & 2.60702 & 2.75076\\
 .8 & 2.25832 & 2.28999 & 2.32698 & 2.37074 & 2.42337 & 2.48792 & 2.56905 & 2.67427 & 2.81662\\
 .9 & 2.32916 & 2.36034 & 2.39676 & 2.43989 & 2.49180 & 2.55554 & 2.63573 & 2.73987 & 2.88097\\
 1  & 2.39792 & 2.42864 & 2.46456 & 2.50713 & 2.55841 & 2.62141 & 2.70076 & 2.80394 & 2.94392\\
 2  & 3.00000 & 3.02812 & 3.06111 & 3.10036 & 3.14787 & 3.20656 & 3.28096 & 3.37843 & 3.51188\\
 3  & 3.50000 & 3.52713 & 3.55903 & 3.59705 & 3.64318 & 3.70031 & 3.77297 & 3.86853 & 4.00000\\
 4  & 3.93700 & 3.96382 & 3.99537 & 4.03303 & 4.07877 & 4.13552 & 4.20784 & 4.30317 & 4.43471\\
 5  & 4.33013 & 4.35695 & 4.38854 & 4.42627 & 4.47214 & 4.52911 & 4.60179 & 4.69776 & 4.83046\\
 6  & 4.69042 & 4.71744 & 4.74927 & 4.78731 & 4.83359 & 4.89112 & 4.96458 & 5.06169 & 5.19615\\
 7  & 5.02494 & 5.05227 & 5.08447 & 5.12298 & 5.16984 & 5.22813 & 5.30261 & 5.40115 & 5.53775\\
 8  & 5.33854 & 5.36625 & 5.39890 & 5.43796 & 5.48552 & 5.54469 & 5.62034 & 5.72050 & 5.85947\\
 9  & 5.63471 & 5.66284 & 5.69600 & 5.73567 & 5.78399 & 5.84413 & 5.92105 & 6.02294 & 6.16441\\
 10 & 5.91608 & 5.94466 & 5.97836 & 6.01868 & 6.06780 & 6.12896 & 6.20721 & 6.31091 & 6.45497\\
 11 & 6.18466 & 6.21371 & 6.24796 & 6.28896 & 6.33891 & 6.40112 & 6.48074 & 6.58630 & 6.73300\\
 12 & 6.44205 & 6.47158 & 6.50641 & 6.54810 & 6.59890 & 6.66218 & 6.74319 & 6.85062 & 7.00000\\
 13 & 6.68954 & 6.71956 & 6.75497 & 6.79736 & 6.84902 & 6.91338 & 6.99580 & 7.10512 & 7.25718\\
 14 & 6.92820 & 6.95872 & 6.99471 & 7.03780 & 7.09033 & 7.15578 & 7.23960 & 7.35082 & 7.50555\\
 15 & 7.15891 & 7.18992 & 7.22649 & 7.27029 & 7.32369 & 7.39022 & 7.47545 & 7.58856 & 7.74597\\
 16 & 7.38241 & 7.41391 & 7.45107 & 7.49558 & 7.54983 & 7.61746 & 7.70409 & 7.81907 & 7.97914\\
 17 & 7.59934 & 7.63134 & 7.66908 & 7.71429 & 7.76940 & 7.83811 & 7.92613 & 8.04299 & 8.20569\\
 18 & 7.81025 & 7.84273 & 7.88106 & 7.92696 & 7.98294 & 8.05271 & 8.14212 & 8.26083 & 8.42615\\
 19 & 8.01561 & 8.04858 & 8.08748 & 8.13408 & 8.19090 & 8.26174 & 8.35253 & 8.47308 & 8.64099\\
 20 & 8.21584 & 8.24929 & 8.28877 & 8.33605 & 8.39372 & 8.46562 & 8.55776 & 8.68014 & 8.85061\\
 \noalign{\smallskip}\hline
\end{tabular}
\end{table}

\textbf{Acknowledgements} I acknowledge the support and encouragement of my family. 



\end{document}